# A genome-wide survey of genetic variation in gorillas using reduced representation sequencing

Aylwyn Scally[1,2]*, Bryndis Yngvadottir[1,3]*, Yali Xue[1], Qasim Ayub[1], Richard Durbin[1], Chris Tyler-Smith[1]

*Equal contribution

[1]The Wellcome Trust Sanger Institute, Wellcome Trust Genome Campus, Hinxton, Cambs. CB10 1SA, UK.
[2]Department of Genetics, University of Cambridge, Cambridge, CB2 3EH, UK
[3]Division of Biological Anthropology, University of Cambridge, Cambridge, CB2 1QH, UK

Keywords: Gorilla; western lowland; eastern lowland; beringei; graueri; genome; sequencing; genetic variation; genetic diversity; heterozygosity; inbreeding; allele sharing; population differentiation; population substructure; demography; centromere; telomere; major histocompatibility locus; MHC; HLA; selective sweep; conservation

**Abstract**
All non-human great apes are endangered in the wild, and it is therefore important to gain an understanding of their demography and genetic diversity. To date, however, genetic studies within these species have largely been confined to mitochondrial DNA and a small number of other loci. Here, we present a genome-wide survey of genetic variation in gorillas using a reduced representation sequencing approach, focusing on the two lowland subspecies. We identify 3,274,491 polymorphic sites in 14 individuals: 12 western lowland gorillas (*Gorilla gorilla gorilla*) and 2 eastern lowland gorillas (*Gorilla beringei graueri*). We find that the two species are genetically distinct, based on levels of heterozygosity and patterns of allele sharing. Focusing on the western lowland population, we observe evidence for population substructure, and a deficit of rare genetic variants suggesting a recent episode of population contraction. In western lowland gorillas, there is an elevation of variation towards telomeres and centromeres on the chromosomal scale. On a finer scale, we find substantial variation in genetic diversity, including a marked reduction close to the major histocompatibility locus, perhaps indicative of recent strong selection there. These findings suggest that despite their maintaining an overall level of genetic diversity equal to or greater than that of humans, population decline, perhaps associated with disease, has been a significant factor in recent and long-term pressures on wild gorilla populations.

**Introduction**
Two species of gorilla are currently recognized, based on morphology and geographical distribution within equatorial Africa: western gorillas (*Gorilla gorilla*), in tropical forest north of the Congo river on the west side of the continent, and eastern gorillas (*Gorilla beringei*) ~1000 km to the east, extending to the slopes of the central African mountains [1,2]. Estimates of when the two species diverged have varied, but recent calculations based on genomic data suggest an initial divergence time of at least 500,000 years ago [3] with gene flow continuing until as recently as 80,000 years ago [4] [5]. Both species are further classified into two subspecies: within the western species, western lowland gorillas (*Gorilla gorilla gorilla*) and Cross River gorillas (*Gorilla gorilla diehli*); within the eastern gorillas, mountain gorillas (*Gorilla beringei beringei*) and eastern lowland gorillas (*Gorilla beringei graueri*).

Western gorillas are classified as critically endangered and eastern gorillas as endangered on the IUCN Red List of Threatened Species [6]. Accurate census statistics are difficult to obtain, but current estimates indicate that wild population levels are rapidly declining, and gorillas of both species generally live in small isolated populations. The major causes of gorilla population decline are thought to be habitat loss, hunting for the bushmeat trade and outbreaks of the Ebola virus [7], and habitat fragmentation in particular has been proposed to have played a role in structuring gorilla genetic diversity [8,9]. In light of these pressures, an understanding of present day genetic variation in gorilla populations, and of their demography, is relevant for conservation efforts. Genetic information is increasingly used, in combination with morphological and ecological data, to classify wild populations and to define (and redefine) taxonomic units for conservation. On longer timescales, understanding the association between past population bottlenecks and inferred climatic or habitat changes may provide valuable insights into the long-term forces affecting gorilla survival [10].

To date, most population-scale studies of genetic variation in gorillas have relied on mtDNA [8,11,12,13] and a limited number of autosomal loci [4]. However, ancestry at individual loci is sensitive to sampling variance, and loci under selection such as mtDNA are susceptible to introgression of haplotypes between populations [14]. Therefore such loci may not be representative of genetic diversity and ancestry across the genome.

Here we present a genome-wide survey of gorilla genetic variation, using a reduced representation sequencing approach to analyse polymorphic loci in twelve western lowland and two eastern lowland gorillas. Using over 3 million SNPs across multiple samples, the largest catalogue of gorilla genetic polymorphism presented to date, we consider the demography of western lowland gorillas, comparing levels of heterozygosity in different individuals and studying the genome-wide distribution of polymorphism and the frequency spectrum of variant alleles. We also use principal component analysis (PCA) and haplotype clustering methods to identify structure within the western lowland population.

**Samples**

We analysed DNA from 14 gorilla individuals (Table 1): 12 western lowland and two eastern lowland gorillas. All gorillas were living in zoos at the time of sampling, and all but three wild-born individuals (Snowflake, Guy and Mukisi) were also born in captivity. For 12 of the individuals, some information about geographical origins was obtained by tracing information about the capture locations of their wild-born ancestors, based on their available family trees (Text S1).

**Sequencing**

We created reduced representation libraries for genome sequencing to explore genetic variation in the 14 gorillas. This method [15,16] reduces the complexity of the genome substantially while ensuring that independently constructed libraries contain mainly identical regions from different individuals dispersed across the genome. Sequencing yielded a combined total of 3,749,197,450 reads of length 100 bp passing quality checking across all 14 individuals. Our chosen *Alu*I target fragments were 150-250 bp in length, estimated to represent 13% of the gorilla genome and to have a substantial reduction in repetitive element content within the chosen fragment size range. After alignment to the gorilla reference genome (gorGor3, Ensembl Release 57) we found the majority of target *Alu*I fragment sites in each individual were covered by a minimum of four reads (Figure S1). An average of 90.3% of reads mapped to the reference, resulting in a total of 91,122,638 bases covered by at least 10 reads in 10 or more individuals, and a total of 3,274,491 SNPs passing coverage thresholds of greater than 10x and less than 100x in at least one gorilla.

One potential concern for studies using this approach is allele dropout, in which haplotypes associated with a mutation at a restriction site are not observed. The impact of this increases with the length of the restriction site motif and the density of polymorphisms, and has been shown to induce bias in estimates of population genetic parameters when enzymes cutting at 6-base or 8-base motifs are used in species with a high mutation rate and effective population size [17]. However, the enzyme *Alu*I used in this study cuts at a 4-base motif, and even with a 6-base cutting enzyme, at the rate of polymorphism found in gorillas ($\sim 10^{-3}$ per bp) the bias is expected to be negligible [17]. Therefore we do not expect our analyses to be significantly biased by allele dropout.

**Genome-wide analysis**

*Polymorphic sites by individual*

Genotypes were called from the alignments of sequence data to the gorilla reference for each of the 14 individuals, with sites filtered on the basis of mapping quality and depth (Methods). Figure 1 shows the resulting rates of heterozygous and homozygous variant sites (with one or both alleles differing from the reference respectively) in each individual. We observed higher levels of heterozygosity in all western gorillas compared to the two eastern gorillas, consistent with previous studies [3,4,18] and indicating that western lowland gorillas are more genetically diverse and have higher effective population size than their eastern relatives. Conversely, we observe higher rates of homozygous variant sites in eastern gorillas than in the individuals classed as western lowland gorillas. Given that the reference genome was assembled from the western lowland gorilla Kamilah [3], this is consistent with the eastern individuals coming from a separate population. (This also accounts for Kamilah's very low homozygous variant rate here – such sites are essentially due to errors in alignment and variant calling, or in the reference.)

With the exception of BiKira, most of the western individuals have a hom/het ratio (ratio of homozygous to heterozygous variant site rates) ranging from 0.5 to 0.7. In an individual whose chromosomes are exchangeable with those of the reference individual (in other words, where both individuals are drawn from a single freely mating population) this ratio is expected to be 0.5 by symmetry. Any departure from this condition, such as the presence of population structure, will manifest as an increase in the hom/het ratio. Thus the variation observed amongst western lowland gorillas may reflect moderate substructure within the western lowland population.

BiKira is a notable outlier, with more homozygous than heterozygous variants. Although this could be a signal of even stronger population structure, an examination of the distribution of variants along her genome reveals long homozygous tracts, indicative of her being inbred (Figure S2). Such inbreeding must presumably have occurred in the wild, since there is no evidence for it in captivity: BiKira's father and both grandparents on her mother's side were born in the wild. Both her father and her grandfather are recorded as having been captured in Cameroon. Of the three other western lowland individuals with highest hom/het ratios, Snowflake is known to exhibit albinism, which may also be a consequence of inbreeding, although, if so, this is not evident in Figure S2.

*Genome-wide distribution of polymorphism in western lowland gorillas*
We investigated the distribution of polymorphism across the autosomal genome by calling sites simultaneously in multiple western lowland individuals (chromosomes X and Y were excluded due to the mixture of male and female samples). Sites were filtered by base quality, read mapping quality and depth of coverage (see Methods), and we also excluded sites for which there was apparent evidence for more than two alleles across all individuals. Although twelve western lowland samples were sequenced, Effie, Kaja and Snowflake were excluded from this comparison because their low depths of sequence coverage would have limited the number of sites callable across all individuals, given the calling thresholds used. Based on these calls, we recorded the resulting density of segregating sites (sites at which two alleles are present across all individuals) in 1 Mbp windows, shown in Figure 2A.

We observe enrichment of polymorphic sites towards both centromeres and telomeres, shown more clearly in Figure 2B and similar to the signal seen in an equivalent analysis of polymorphism in humans [19]. Some regions stand out as particularly dense in polymorphisms, notably a pericentromeric cluster on chromosome 9 (which may reflect mismapping to a complex repetitive region) and several regions on the short arm of chromosome 8 (enriched for structural variation in humans). Relative to the genome-wide average, we observe intermediate to high levels of diversity in the major histocompatibility complex (MHC) region on chromosome 6 (Figure S4), but the increase is markedly less prominent than is found at this region in chimpanzees (Figure S5) [20] and humans [19]. Furthermore, in the region immediately upstream there is a clear reduction in genetic diversity (Figure S4), with a profile extending over 5 Mbp centred around chr6:27,500,000 which is suggestive of a recent selective sweep in western lowland gorillas at this locus. Several genes are located there, including some associated with immune response: the *BTN2* subfamily, which forms part of the immunoglobulin superfamily, and *PRSS16*, which encodes for thymus-specific serine protease, thought to be involved in the prevention of auto-immunity [21].

*Allele frequency spectrum*
The allele frequency spectrum (AFS) within a population provides insight into demographic and evolutionary processes. We obtained conditional allele frequency spectra by identifying heterozygous sites within one individual (taking each individual in turn) and estimating the non-reference allele frequency across all samples at each of those sites (Methods). Figure 3 shows the resulting mean conditional AFS for nine western lowland gorillas (excluding the three lowest-coverage samples as above), and comparable samples from three human populations whose ancestry derives from Africa, Asia and Europe [22]. In a population whose effective size has remained constant, the theoretical expectation for such a conditional AFS is a straight line of constant negative slope [23], shown by the dashed line in Figure 3. Compared to this, western lowland gorillas show a deficit of rare alleles, consistent with their having undergone genetic drift due to a bottleneck or other reduction in effective population size during their demographic history. The similar signal in non-African human populations has been attributed to population contraction associated with the out-of-Africa event [24]. By contrast to the gorillas and the non-African humans, the African YRI population in Figure 3 shows an excess of rare alleles, consistent with population expansion and again similar to the signal seen in other African human data [24].

*Population structure*
We carried out a principal component analysis (PCA) to evaluate population structure in gorillas, based on genotypes called in each individual and filtered by quality and depth as above (Methods). The results support a division between eastern and western gorillas (Figure 4A), in which the first principal component represents the east-west species distinction and the second principal component captures variation primarily within the eastern population, albeit for two samples only. An ADMIXTURE analysis on the same samples also supported this division, finding an optimal separation into two clusters which segregate the eastern and western gorillas (Figure S5).

To further explore structure within western lowland gorillas, we carried out PCA restricted to the western samples only. The resulting decomposition revealed substructure within western lowland gorillas manifesting as clustering in the first principal component (Figure 4B), and separating Ruby and Guy from the others. The second principal component separates the albino gorilla Snowflake from the others; however given the number of samples this separation may not be significant.

**Discussion**

We present a catalogue of gorilla population genetic diversity across the dispersed subset of the genome targeted by our reduced representation sequencing strategy, the largest to date in terms of the number of loci studied. Based on these data, we have explored demographic differences between and within gorilla populations, and carried out an initial genome-wide survey of western lowland gorilla genetic polymorphism.

*Gorilla demography*

Our analyses of heterozygous and homozygous variant rates (Figure 1) and of PCA decomposition (Figure 4) demonstrate a clear division between individuals from the western and eastern species. Both analyses also reveal evidence for further population structure within the western lowland gorillas. Such structure has been suggested previously in genetic studies of isolated loci [4,25]. Earlier observations of morphological features also suggested the existence of four "demes" within the western lowland gorilla range based on clinal variations in skull size [26]. Population substructure is often attributed to an isolation by distance effect, and this has been proposed in the case of western lowland gorillas [9], but alternative causes have also been suggested [4]. For example, a recent study of gorilla dental data found that gorilla population structure correlates more significantly with altitudinal variation than with geographic distance, suggesting that isolation by distance may not provide a complete explanation for population dispersal patterns in gorillas [2].

*Genetic polymorphism in western lowland gorillas*

Within western lowland gorillas, we observed enrichment in genetic diversity near telomeres and centromeres, similar to that seen in humans [19]. However the detailed pattern of variation differs on the 1 Mbp scale between gorillas, humans and chimpanzees, and in particular we observed only a modest increase in gorilla genetic diversity at the major histocompatibility complex (MHC) locus on chromosome 6 relative to the genome wide average. By comparison with chimpanzees, and particularly with humans, there is a strong signal of increased diversity in this region (Figure S5; [19]). A previous study of a small number of MHC coding loci found similar levels of diversity in both eastern and western gorilla species, but did not compare to other regions of the genome [27], and given that eastern gorillas are known to be less diverse genome wide [3], this is not inconsistent with our finding.. Genes encoded by the MHC play a key role in antigen recognition, and it has been suggested that increased MHC heterozygosity provides a fitness advantage for individuals, leading to high genetic diversity within populations [28,29]. However the degree to which this is borne out in wild populations remains unclear, as does the precise mechanism by which selection might favour MHC diversity [30]. In western lowland gorillas, diversity upstream of the MHC locus is in fact lower than the genome wide average, and has a profile consistent with there having been a recent selective sweep there (although more sensitive haploytype-based analyses would be required to explore this hypothesis). We note also that notwithstanding the increased diversity in western chimpanzees across the MHC locus as a whole (Figure S6), evidence of a selective sweep at the MHC Class I region, manifesting as reduced allelic diversity there, has been reported previously in this subspecies [31]

Our analysis of the allele frequency spectrum in western lowland gorillas revealed a deficit of low-frequency alleles relative to expectation under a model of constant population size. A similar signal in non-African humans has been attributed to the reduction in population size during the exodus from Africa, evident despite their subsequent exponential growth in population size. Although there is no evidence for western lowland gorillas having undergone large-scale migration in a similar manner, it may be that they have experienced an episode of population collapse (perhaps associated with a severe disease epidemic), followed by a recovery period of population growth. It remains the case that western lowland gorillas show a higher overall level of genetic diversity than humans.

*Future prospects*

To date, our understanding of gorilla demography and evolution has been limited both by the number of samples and the number of loci studied. The genetic distinction between eastern and western gorillas is clear, but whereas most field studies have been carried out on eastern gorillas [32,33], genetic studies have been largely confined to western gorillas. This is due in part to the difficulty of obtaining high-quality DNA from individuals in the wild, and to the fact that the zoo population is predominantly comprised of western lowland individuals. A better understanding of gorilla population history and genetics will require more data from wild-born individuals, combined with information about their geographical origin. As sequencing costs continue to decrease, we look forward to the prospect of whole genome sequences from multiple individuals and from all known subpopulations. Future studies collecting such data will be able to address more complex demographic questions, such as the sizes of subpopulations, their degree of substructure, and rates of gene flow. In addition, such data can play an important role in the management and conservation of a species. For example, the identification of substructure within western lowland gorillas can inform and may influence breeding strategies. These efforts are particularly important in light of the endangered status and increasing isolation of gorilla populations caused by habitat fragmentation and rapid decline [6].

## Methods

*Samples*

One sample (EB(JC)) was purchased from the European Collection of Cell Cultures (ECACC, Porton Down, Salisbury, UK). The remaining samples from 13 gorillas (Table 1) have been obtained over several years, with the help of carers and veterinarians working at the zoos where the gorillas reside, and our colleagues (see Acknowledgements). For some gorillas, blood samples were provided, while from others high quality DNA was already available. Blood samples were collected when the gorillas were anesthetized during standard veterinary procedures.

*Reduced representation sequencing*

Approximately 1 µg of gorilla DNA was digested overnight with *Alu*I restriction enzyme and the digested fragments were purified with Qiagen PCR cleanup columns (Hilden, Germany) and re-suspended in 36 µl of TE buffer. Fragment sizes ranging from 150-250 bp were used for library construction. Initially the samples were sized on the Pippin Prep (Sage Science, Beverly, Massachusetts, USA) using 2% cassettes containing ethidium bromide and marker B as the reference ladder. After sizing, the samples were run through a Qiagen quick column and an Agilent Bioanalyzer (Clara, California, USA) to check if the sizing was successful, following which the samples went onto standard library preparation employing NEBNext® DNA Library Prep Master Mix Set for Illumina® (New England Biolabs Ipswich, Massachusetts, USA). There were three main stages to library preparation; end repair, A-tailing and adapter ligation. The adapters were supplied by Integrated DNA Technologies (Coralville, Iowa, USA). After both end repair and A-tailing, the samples were purified using Qiagen quick columns. After adapter ligation, Agencourt AMPure XP beads (Beckman Coulter**,** High Wycombe, UK) were used to purify the samples and remove adapter duplex. Each library preparation was checked on an Agilent Bioanalyzer and amplified using 8 cycles of PCR with PE1.0 and PE2.0 as primers and 2X Kapa HiFi hotstart ready mix. Agencourt AMPure XP beads were used to purify the samples and remove any primers remaining after the PCR. Each library was sequenced on a single lane of an Illumina HiSeq (100 bp paired end reads).

Primary sequencing data generated for this study are available in public sequence databases under accession number ERP000748.

*Alignment and variant calling*

Sequence reads were aligned to the gorilla assembly (gorGor3, Ensembl Release 63, June 2011) using the BWA program [34]. Genotypes were called using SAMtools [35] and filters based on minimum read base quality (20), minimum r.m.s. mapping quality of the covering reads (25), minimum (10) and maximum (100) coverage depth were applied. Sites with apparent evidence for more than two alleles across all samples were also excluded. Variant calls in VCF format are available to download from ftp.sanger.ac.uk/pub/ylx/gorilla/rr_vcf .

*Validation*
We investigated the concordance between variants called in Kamilah using the reduced representation (RR) data obtained in this study and a previous high-coverage whole-genome (WG) data set [3]. Applying the same algorithm to both sets of data, we found that 96.2% of heterozygous calls in the RR data are also found in calls made from the WG data set. Also, 85% of WG heterozygous sites within regions that were covered at 10x or more in the RR data were also called as heterozygous in the RR data, with a further 4% called as homozygous non-reference variants.

Of 12 private homozygous variants that were selected for validation by Sanger sequencing, primers were designed for nine, PCR products obtained for six and sequence for five, of which three were validated as homozygous variants specific to that individual.

*Genome-wide polymorphism*
Sites were called across the nine highest-depth gorilla samples (i.e. excluding Effie, Kaja and Snowflake) using the filtering thresholds as above, and each site was classified as segregating or non-segregating, according to number of alleles present across all samples. Based on these calls the density of segregating sites was binned in 1 Mbp windows across the gorilla genome.

Published sequence data (alignment files) for ten western chimpanzees (*Pan troglodytes verus*) were downloaded from http://panmap.uchicago.edu/ and processed by calling segregating sites across all samples. Filtering was as above except that due to the lower mean coverage of this data a minimum depth threshold of 7 reads was used.

*Allele frequency spectra*
Allele frequency spectra (AFS) in the western lowland gorilla samples were calculated as follows. Taking one individual, heterozygous sites in that individual were identified, based on the genotypes called above. At each of these sites, the non-reference allele frequency was estimated across all samples using SAMtools, and a conditional AFS over all sites constructed from these frequencies. Conditioning on each individual in turn generated multiple AFS, and an average conditional AFS was then calculated by taking the mean over all spectra at each frequency. This procedure was carried out on the nine highest depth western lowland gorillas in our study (BiKira, EB(JC), Fubu, Guy, Kamilah, Kesho, Matadi, Murphy and Ruby). For comparison, we also calculated conditional AFS using published data from the 1000 Genomes Project for three human populations [22]: YRI, CEU and CHB. Eleven unrelated samples were used in each case: (CEU: NA06984, NA06986, NA06989, NA06994, NA07000, NA07037, NA07048, NA07051, NA07056, NA07347, NA10847, NA11830; CHB: NA18530, NA18534, NA18536, NA18543, NA18544, NA18546, NA18525, NA18526, NA18527, NA18528, NA18532, NA18535; YRI: NA18486, NA18487, NA18489, NA18498, NA18499, NA18501, NA18502, NA18504, NA18505, NA18507, NA18508, NA18511), five males and six females, with genotypes for each taken from the VCF files at http://ftp.1000genomes.ebi.ac.uk/vol1/ftp/phase1/analysis_results/integrated_call_sets/.

*Population structure*
For the population structure analysis we filtered our SNP data set by removing SNPs that were in Hardy-Weinberg disequilibrium in western lowland gorillas and only using sites that passed our coverage filter (10-100x) and were called in at least 4 gorillas. We performed principal component analysis (PCA) implemented in EIGENSTRAT [36] on our filtered dataset.

We used the program ADMIXTURE [37] to perform a maximum likelihood estimation of individual ancestries in the gorillas. Each ADMIXTURE analysis requires a hypothesized number of ancestral populations (K) and assigns individuals to these populations without using any pre-assigned population labels. We ran this on the filtered dataset described above, with one additional filter: we pruned the data to remove SNPs that were in detectable LD ($r^2 > 0.1$) as recommended for this type of analysis in the ADMIXTURE manual. Cross-validation was used to estimate the optimum number of clusters (K).

**Acknowledgements**
We are very grateful to all who provided us with samples from the gorillas on which this research is based upon. In particular, we would like to thank Ilona Furrokh, David A. Field, Andy Hartley and Nick Lindsay at ZSL London Zoo for samples from Effie and Kesho, Michael Griffith, John Fisher and Julie Mansell at Belfast Zoological Gardens for BiKira's sample, Neil Spooner and Jane Hopper at Howletts Wild Animal Park for Fubu's sample, Tracy Cork, John Lewis and Marc Boardman at Chessington Zoo for Kaja's sample, Bridget Fry and Nic Masters at Twycross Zoo for Matadi's sample, Jane Rogers and San Diego Zoo for Kamilah's sample, Peter Galbusera and Antwerp Zoo for Victoria's sample, David Comas and Jaume Bertranpetit from Universitat Pompeu Fabra, Barcelona for Snowflake's sample, Mark A. Jobling from the University of Leicester for Ruby and Guy's samples, and finally K. Leus of the Center for Research and Conservation of the Royal Zoological Society of Antwerp for Mukisi's sample, from which DNA was extracted and provided by Olaf Thalmann and Linda Vigilant. We thank James R. Davis and Robert A. Hess for access to their database and program "Data Viewer for Gorillas in Zoos Worldwide" to enable us to collect information on the gorillas and construct their family trees, and the Sanger Institute DNA Pipelines Core for generating the sequence data. Thanks also to Luca Pagani for helping with the PCA and ADMIXTURE analyses.

**Figures**

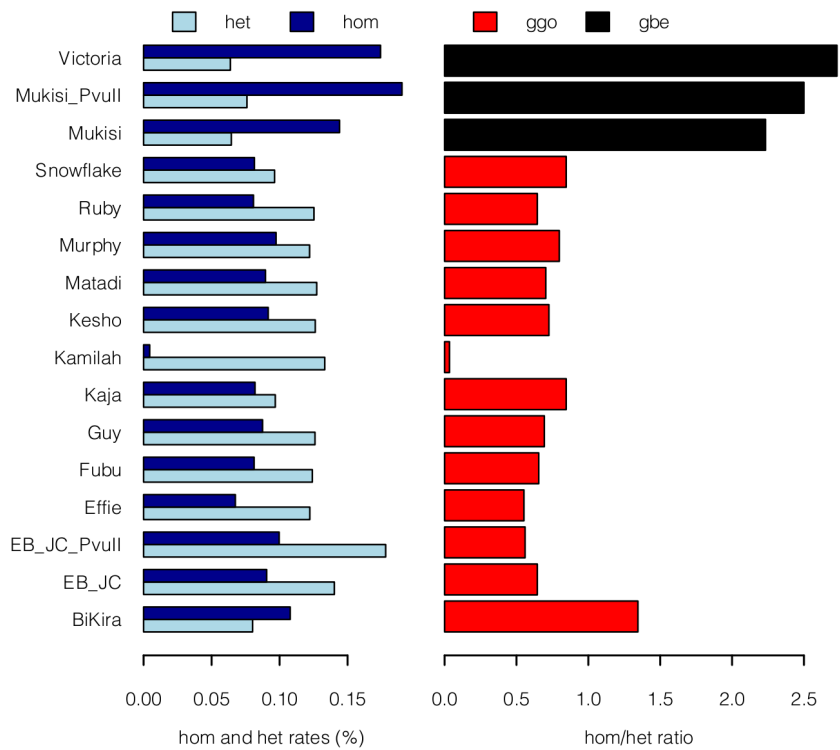

**Figure 1: Rates and ratios of heterozygous and homozygous variants in each of the gorillas sampled.** Rates of heterozygous (light blue) and homozygous (dark blue) variants were called from sequence alignments against the gorilla reference genome and expressed as percentage rates. Note Kamilah's low rate of homozygous variants, due to her providing the DNA from which the reference genome was assembled. The corresponding hom/het ratios (ratios of homozygous to heterozygous variant rates) show that eastern lowland gorillas (black) have higher ratios than western lowland gorillas (red). Additional data for Mukisi and EB(JC) (Mukisi_PvuII and EB_JC_PvuII) were taken from a previously published study [3].

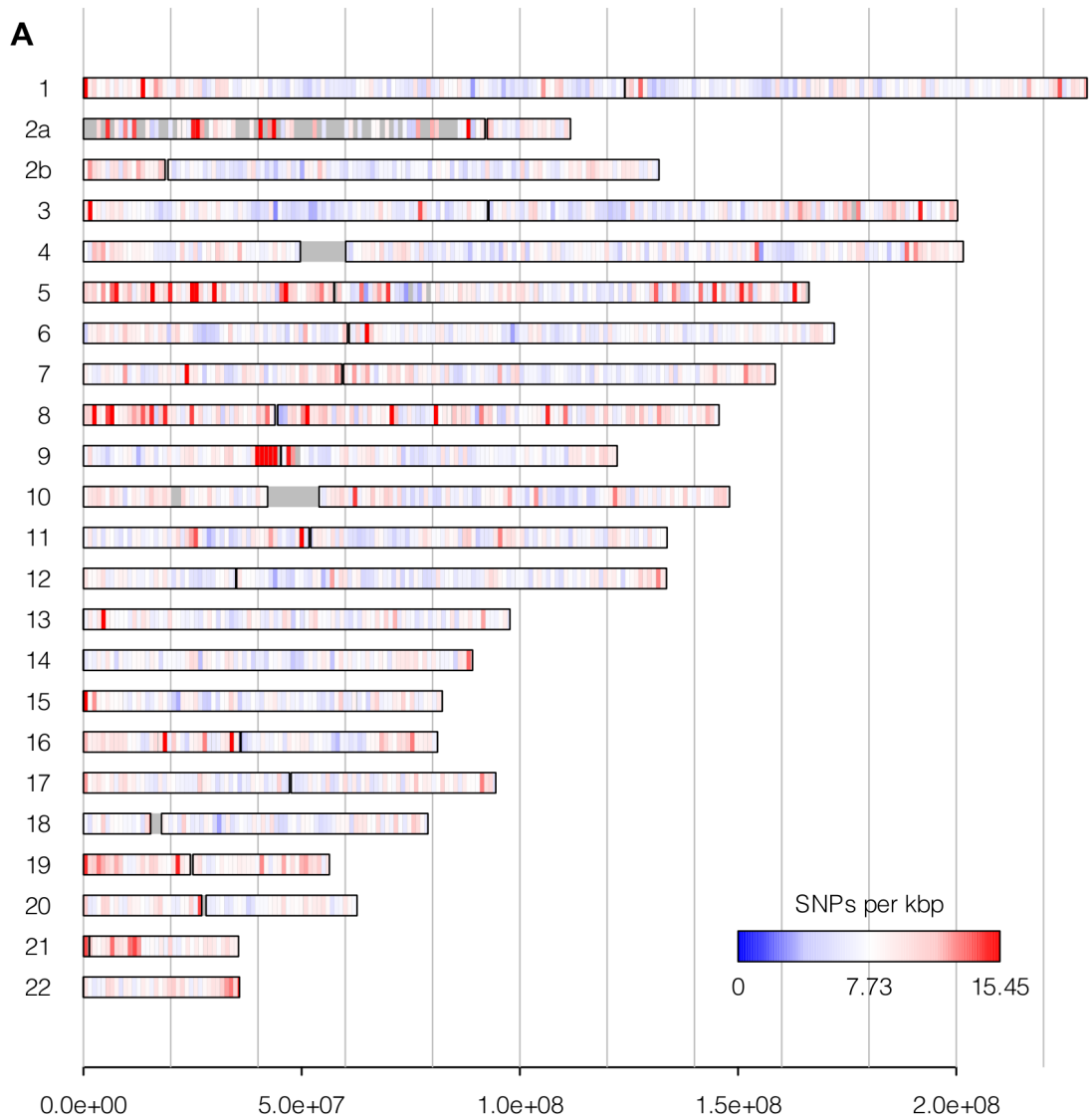

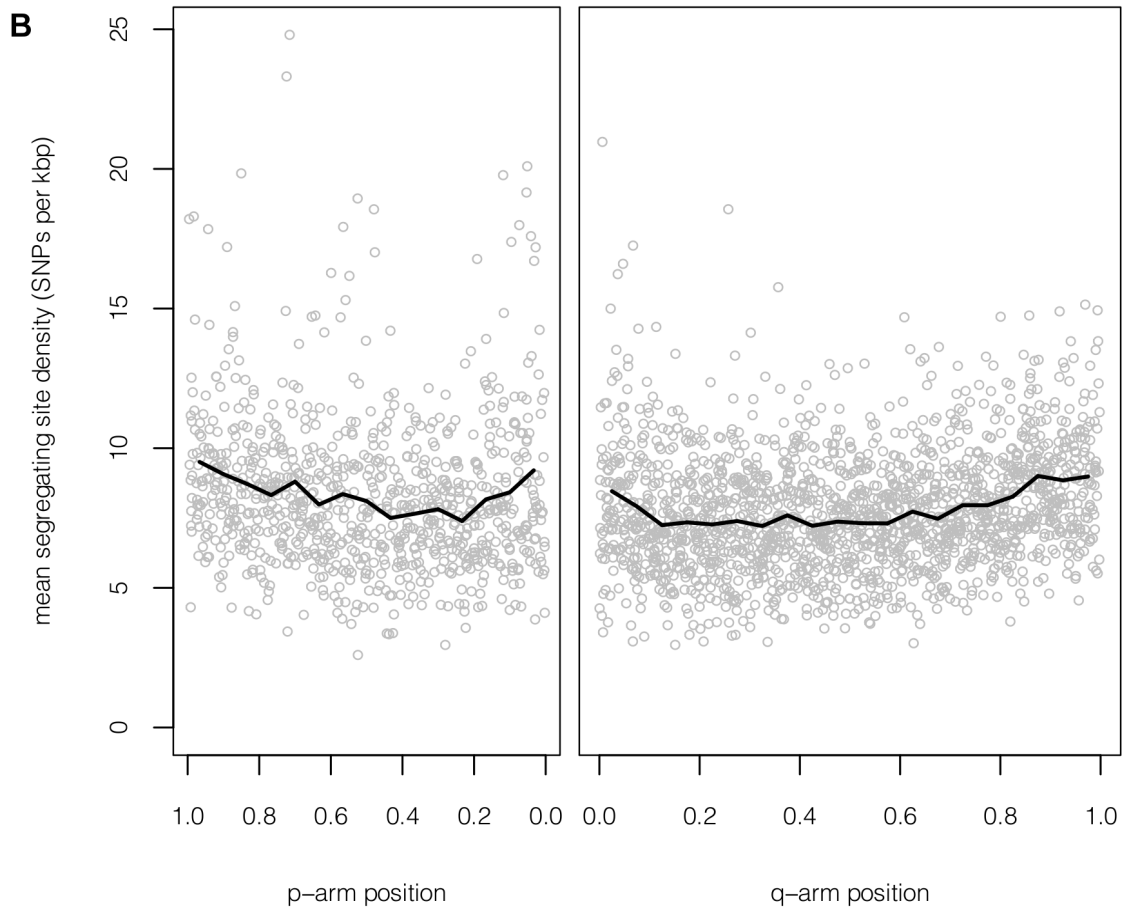

**Figure 2: Segregating sites in nine western lowland gorillas**. **A**. Density of segregating sites in 1Mbp bins. Sites passing quality and depth filtering thresholds in all nine gorillas BiKira, EB(JC), Fubu, Guy, Kamilah, Kesho, Matadi, Murphy and Ruby were binned in 1 Mb bins and the density of segregating sites calculated. Resulting densities are plotted on an ideogram, with the scale expressed as number of sites per kbp. **B.** Profile of mean segregating site density as a function of chromosomal position on both long and short arms, averaged over all chromosomes. Position is normalised by chromosome length, with the centromere at 0.0 and the telomere at 1.0 on both arms. An increase in genetic diversity is evident towards the centromere and telomere on both arms.

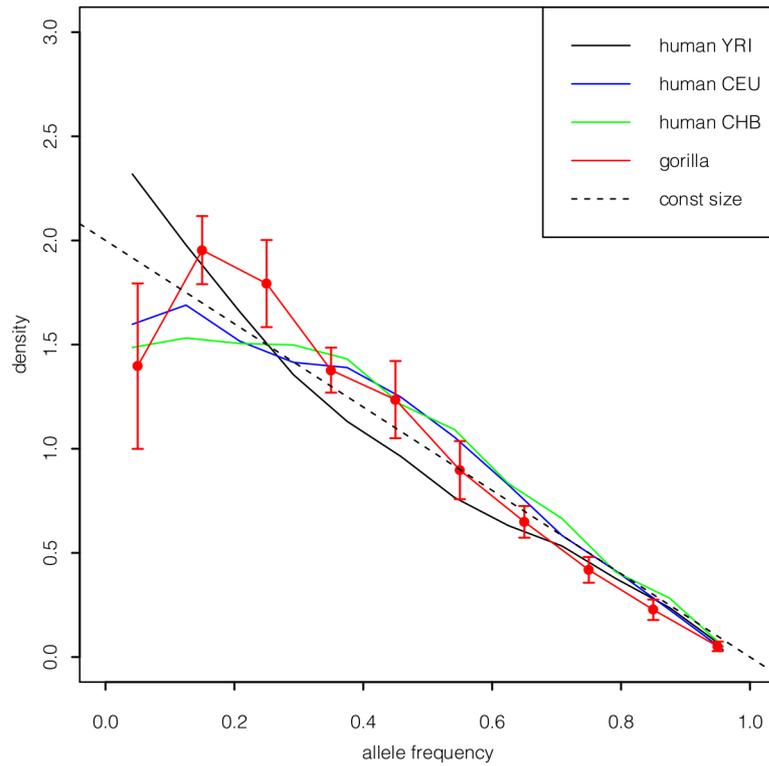

**Figure 3: Allele frequency spectra.** Conditional allele frequency spectra are shown for nine western lowland gorillas (red) and 11 samples from each of three human populations: YRI (black), CEU (blue) and CHB (green). Spectra are conditioned on sites ascertained in one individual (and, for gorilla, averaged over all samples). The dashed line is the theoretical expectation for a constant population size. Error bars for the gorilla samples represent standard deviations.

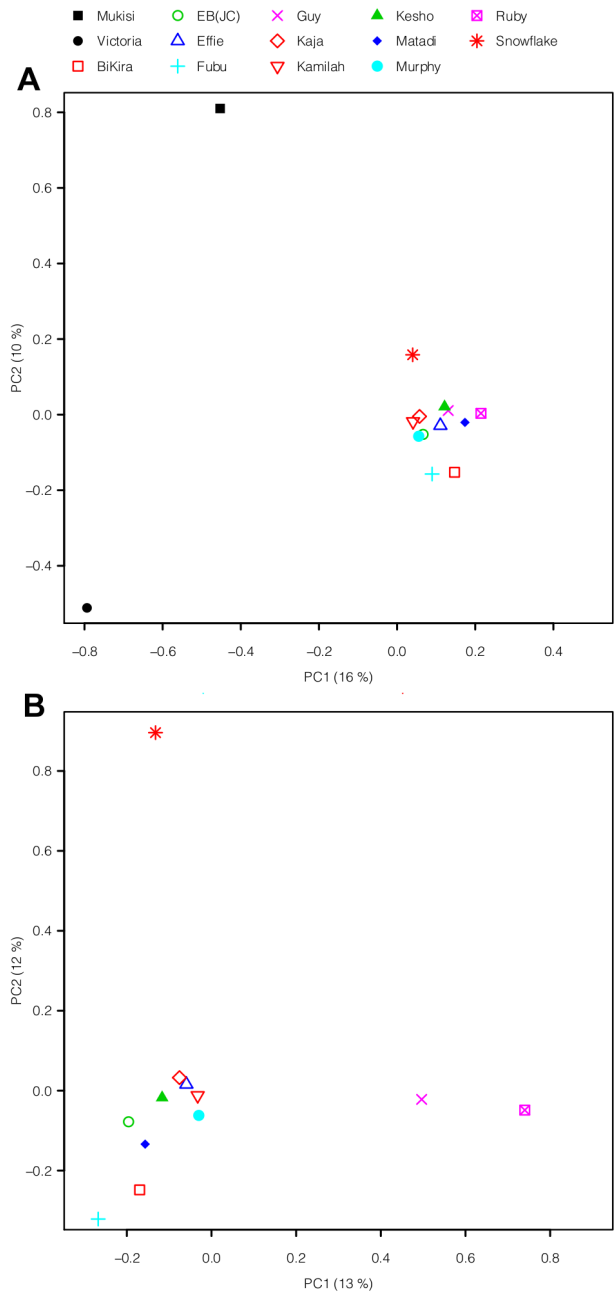

**Figure 4: Principal components analysis (PCA) A**. PCA based on 3,586,308 polymorphic sites in 12 western lowland gorillas and two eastern lowland gorillas. Here, PC1 separates western gorillas from eastern gorillas. **B**. PCA based on 3,263,029 polymorphic sites in the 12 western lowland gorillas only.

**Table 1: Gorillas sampled in this study.**

**Western lowland gorillas**

| Sample name | Studbook ID | Sex | ENA* accession ID |
| --- | --- | --- | --- |
| BiKira | 1352 | F | ERS039715 |
| EB(JC) | N/A | F | ERS-awaited |
| Effie | 1287 | F | ERS039712 |
| Fubu | 1742 | F | ERS039718 |
| Guy | 0005 | M | ERS039708 |
| Kaja | 0700 | F | ERS039711 |
| Kamilah | 0661 | F | ERS014181 |
| Kesho | 1521 | M | ERS039714 |
| Matadi | 1745 | M | ERS039713 |
| Murphy | 0684 | M | ERS040142 |
| Ruby | 1437 | F | ERS039709 |
| Snowflake | 0281 | M | ERS039710 |

**Eastern lowland gorillas**

| Sample name | Studbook ID | Sex | ENA* accession ID |
| --- | --- | --- | --- |
| Mukisi | 9912 | M | ERS039717 |
| Victoria | 9919 | F | ERS040148 |

*ENA: European Nucleotide Archive (http://www.ebi.ac.uk/ena/).

**Supplementary figures**

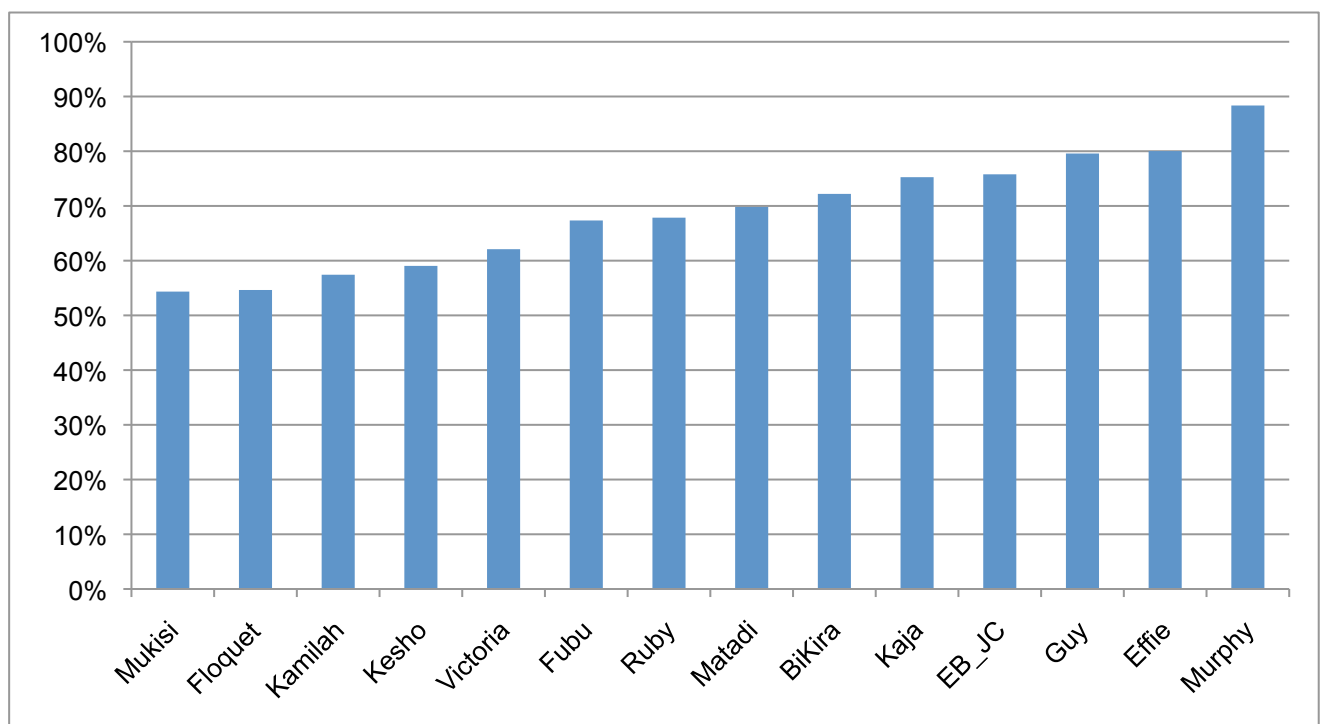

**Figure S1: Percentage of target *Alu*I fragments sites (150-250) bp covered in the 14 gorillas sequenced.**

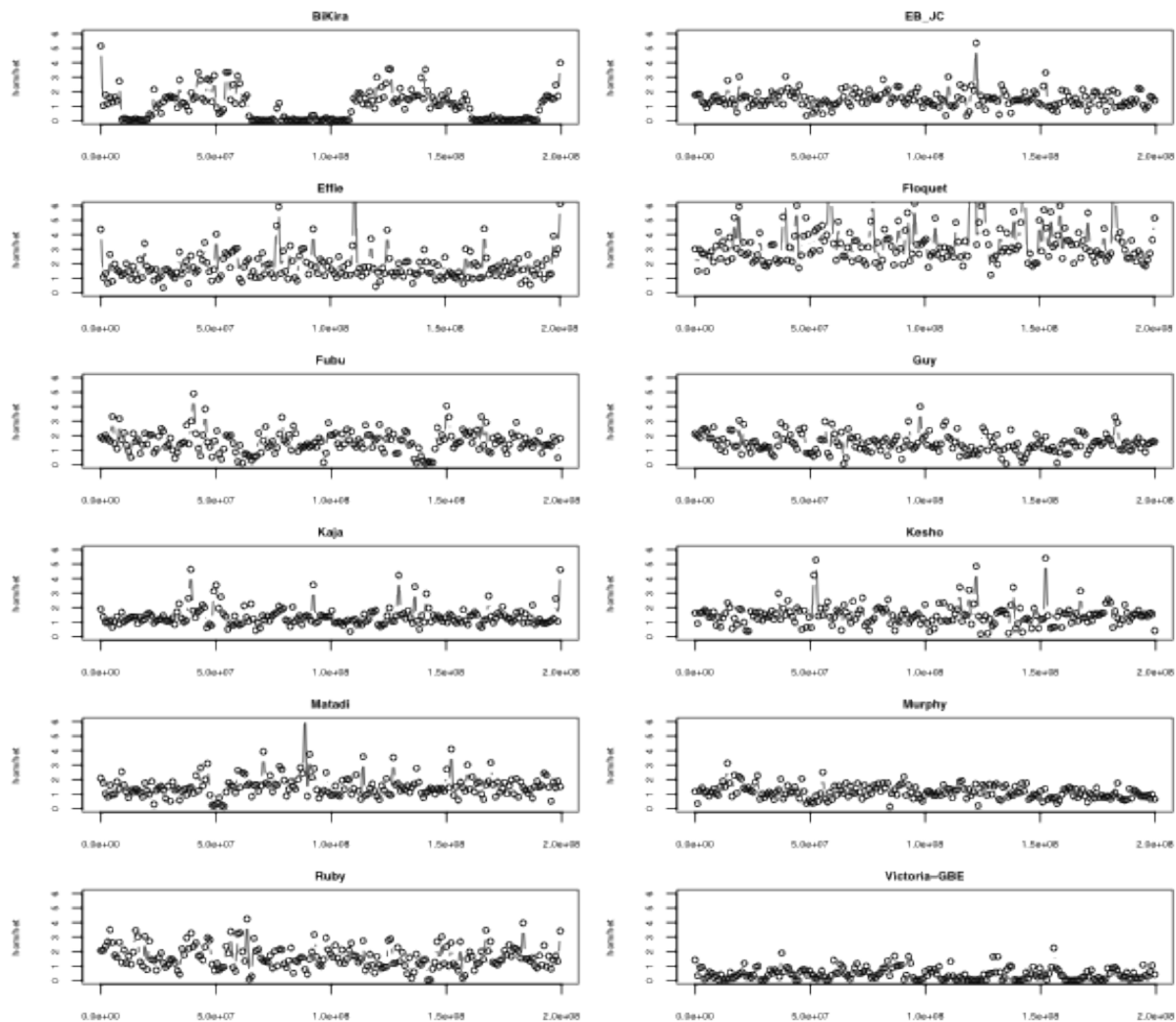

**Figure S2: Ratio of heterozygous to homozygous at variant sites in 1 Mbp bins along Chromosome 3 for 12 gorillas**. Note the extended tracts of homozygosity in BiKira. Other chromosomes (not shown) display a similar pattern.

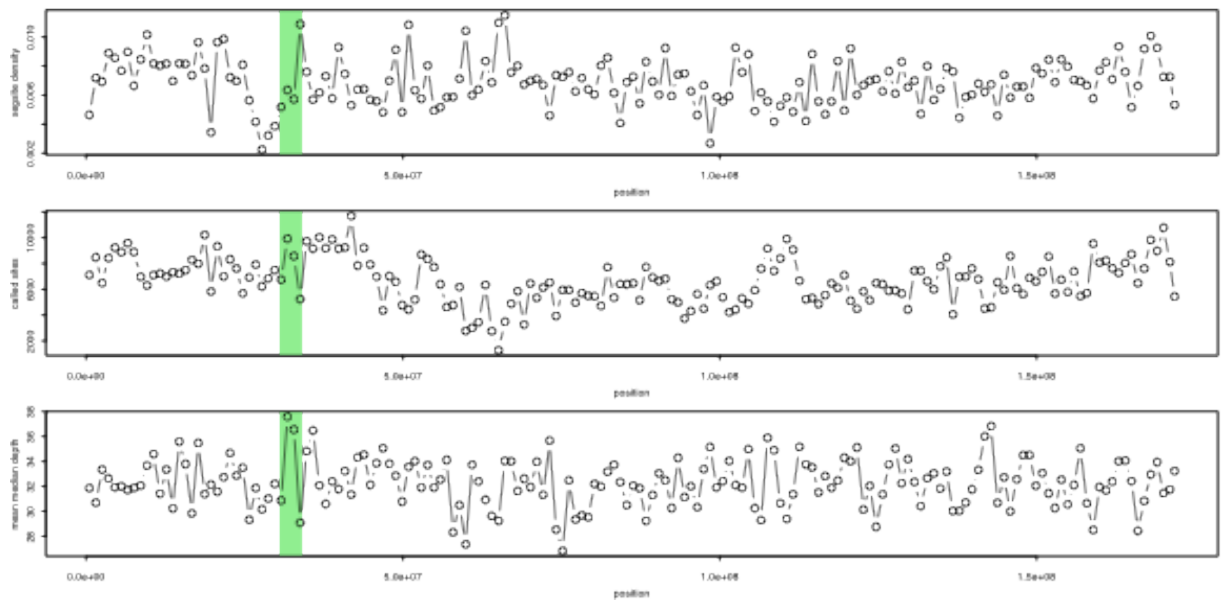

**Figure S3: Western lowland gorilla polymorphism rate, called sites and mean median depth along Chromosome 6**. Each point represents a 1 Mbp bin. The green shaded region indicates the location of the MHC.

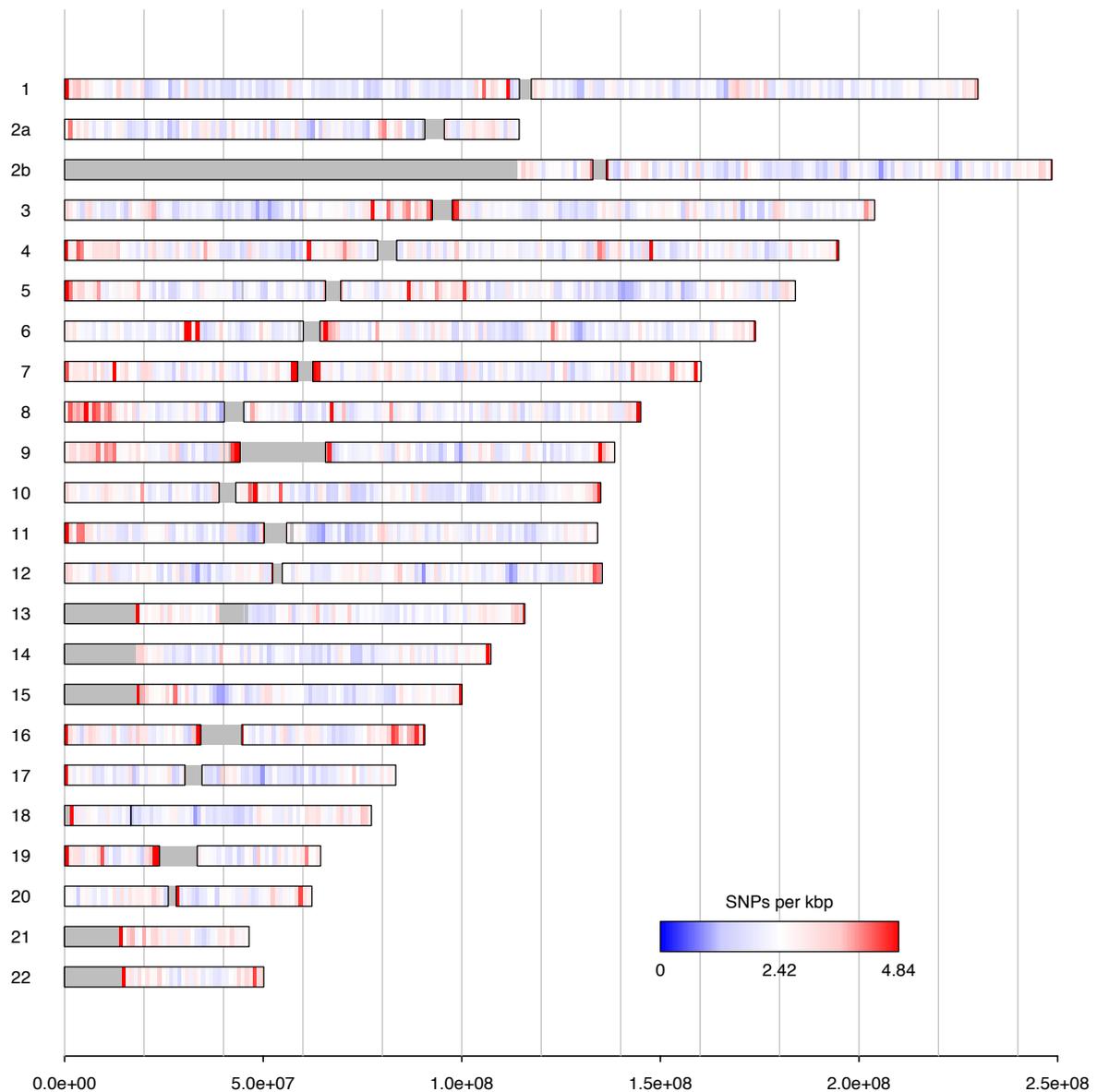

**Figure S4: Density of segregating sites in ten western chimpanzees (*Pan troglodytes verus*)**. Segregating sites are based on genome-wide sequencing data from [20]. Sites passing quality and depth filtering thresholds in all ten samples were binned in 1 Mb bins and the density of segregating sites calculated. Resulting densities are plotted on an ideogram, with the scale expressed as number of sites per kbp. Note the prominent signal of increased diversity at the MHC locus on 6p.

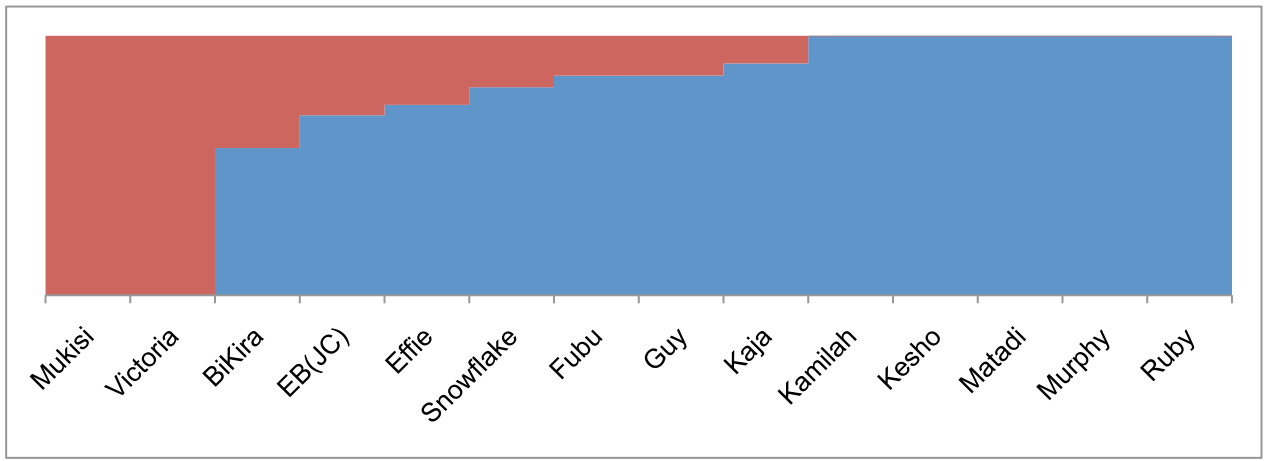

**Figure S5: ADMIXTURE plot of eastern lowland and western lowland gorillas**. Each gorilla is represented by a single vertical bar and the proportion of ancestry is displayed in different colours. For K=2, red represents the eastern component and blue the western component.